# Metallic vanadium disulfide nanosheets as a platform material for multifunctional electrode applications

Qingqing Ji[1+], Cong Li[1,2+], Jingli Wang[3], Jingjing Niu[4], Yue Gong[5], Zhepeng Zhang[1], Qiyi Fang[1,2], Yu Zhang[1,2], Jianping Shi[1,2], Lei Liao[3], Xiaosong Wu[4], Lin Gu[5,6,7*], Zhongfan Liu[1*], Yanfeng Zhang[1,2*]

[1]Center for Nanochemistry (CNC), Beijing National Laboratory for Molecular Sciences, College of Chemistry and Molecular Engineering, Academy for Advanced Interdisciplinary Studies, Peking University, Beijing 100871, People's Republic of China

[2]Department of Materials Science and Engineering, College of Engineering, Peking University, Beijing 100871, People's Republic of China

[3]Department of Physics and Key Laboratory of Artificial Micro- and Nano-structures of Ministry of Education, Wuhan University, Wuhan 430072, People's Republic of China

[4]State Key Laboratory for Artificial Microstructure and Mesoscopic Physics, Peking University, Beijing 100871, People's Republic of China

[5]Beijing National Laboratory for Condensed Matter Physics, Institute of Physics, Chinese Academy of Sciences, Beijing 100190, People's Republic of China

[6]Collaborative Innovation Center of Quantum Matter, Beijing 100190, People's Republic of China

[7]School of Physical Sciences, University of Chinese Academy of Sciences, Beijing 100190, People's Republic of China

* Address correspondence: yanfengzhang@pku.edu.cn; zfliu@pku.edu.cn; l.gu@iphy.ac.cn.

+ These authors contributed equally to this work.

**Abstract** Nano-thick metallic transition metal dichalcogenides such as $VS_2$ are essential building blocks for constructing next-generation electronic and energy-storage applications, as well as for exploring unique physical issues associated with the dimensionality effect. However, such 2D layered materials have yet to be achieved through either mechanical exfoliation or bottom-up synthesis. Herein, we report a facile chemical vapor deposition route for direct production of crystalline $VS_2$ nanosheets with sub-10 nm thicknesses and domain sizes of tens of micrometers. The obtained nanosheets feature spontaneous superlattice periodicities and excellent electrical conductivities ($\sim 3\times 10^3$ S cm$^{-1}$), which has enabled a variety of applications such as contact electrodes for monolayer $MoS_2$ with contact resistances of ~1/4 to that of Ni/Au metals, and as supercapacitor electrodes in aqueous electrolytes showing specific capacitances as high as $8.6\times 10^2$ F g$^{-1}$. This work provides fresh insights into the delicate structure-property relationship and the broad application prospects of such metallic 2D materials.

Metallic transition metal dichalcogenides (MTMDCs) have manifested a wealth of intriguing properties in their bulk states, such as magnetism[1-3], charge density waves[4-7], and superconductivity[8-10], resulting in worldwide attention from condensed-matter physicists over several decades. Recently, there has been renewed research interest in these MTMDCs ($TaS_2$, $NbSe_2$, *etc.*), as they have proven to be ideal systems for exploring collective electronic states down to the two-dimensional (2D) limit[11-14]. More intriguingly, the excellent electrical conductivity and lack of bandgaps in these layered materials also indicate a broad range of potential applications such as transparent electrodes[15] and energy conversion/storage[16], if they can be thinned down to the nanoscale. However, the acclaimed 2D forms remain unattainable and almost unexplored to date, which is in stark contrast to semimetallic graphene[17, 18] and their dichalcogenide analogues such as monolayer $MoS_2$[19, 20], $MoTe_2$[21,22], and $ReS_2$[23,24]. This is mainly attributed to the hysteretic research developments in the batch production of high-quality, large-domain-size MTMDC nanosheets[14, 25], and of paramount importance, with sub-10 nm thickness that approaches the 2D regime[26, 27].

Vanadium disulfide ($VS_2$) is a representative member of the MTMDC family that has a partially-filled energy band at the Fermi level ($E_F$). Being a model $d^1$ system with every V atom having one unpaired *d*-orbital electron[28], this metallic van der Waals (vdW) material is attractive for investigating electronically ordered phases that are susceptible to structural instabilities[29, 30]. More interestingly, it has been predicted by theoretical calculations that $VS_2$ possesses unique dimension-dependent magnetic properties[31], rendering it the first 2D ferromagnet[32] that holds promise for next-generation spintronic applications. However, the intricate V-S phase diagram contains various intermediate compounds ($V_3S_4$, $V_5S_8$, $VS_2$, *etc.*) that can be easily transformed into each other by thermal annealing[33]. This has made achieving pure-phase $VS_2$ for accurate determination of its structure-property relationship a great challenge even in the bulk state[30], not to mention for the 2D counterparts.

Conventionally, bulk $VS_2$ crystals are often prepared using chemical vapor transport techniques[34-36], which are not only time-consuming but also yield nonstoichiometric samples with excess V atoms within the vdW gaps (the space between two neighboring VS$_2$ layers). This deviation from the ideal vdW layered structure has made top-down exfoliation methods ineffective for the preparation of nano-thick VS$_2$ flakes. On the other hand, current bottom-up synthetic strategies, either by hydrothermal methods[37] or chemical vapor deposition[38] (CVD), are still incapable of mass-producing VS$_2$ nanosheets that simultaneously possess sub-10 nm thicknesses and 100 μm domain sizes. This has not only retarded the exploration of their thickness-dependent electronic/magnetic properties, but also greatly impeded their practical applications from being realized, although previous experimental explorations have demonstrated a great promise for electrochemical energy storage[37] and hydrogen evolution applications[38].

To tackle the above issues, we herein report an ambient-pressure CVD growth of VS$_2$ nanosheets with sub-10 nm thickness and lateral sizes of up to tens of micrometers, which has offered us a great opportunity to explore the detailed atomic structures and electronic properties of the material close to the 2D limit. Interestingly, the VS$_2$ nanosheets, either vertically aligned or laid down on SiO$_2$/Si, can be facilely transferred to arbitrary substrates even without the aid of polymer supports. This transferability, combined with the scalability of the CVD techniques, has enabled the VS$_2$ nanosheets to be used as promising electrode materials for supercapacitors and monolayer MoS$_2$ electronics, exhibiting remarkable performances in association with their metallic properties and ultrahigh density of states (DOS) at $E_F$. The findings presented in this work are believed to pave the way for the property investigation and application development with this metallic 2D material.

**Results**

**CVD growth of the VS$_2$ nanosheets.** The VS$_2$ samples were grown by a facile CVD method under a mixed Ar/H$_2$ gas flow, with the substrates (such as SiO$_2$/Si) kept at ~600 °C, and solid VCl$_3$ and sulfur used as precursors (Figure 1a). It was found that, lowering the evaporation temperature of VCl$_3$ to 275–300 °C was the key to growing ultrathin VS$_2$, along with the location optimization of SiO$_2$/Si substrates with respect to the VCl$_3$ precursor (see Supplementary Figure S1, S2 for details). Typically, three kinds of VS$_2$ flakes were found on SiO$_2$/Si substrates: thick hexagonal VS$_2$ (>100 nm), and ultrathin VS$_2$ (<10 nm) that was either vertically grown or laid down on the substrates (Figure 1b). The unique half-hexagon domain shapes of the ultrathin VS$_2$ can be understood as toppling of the vertically-aligned VS$_2$ sheets whose longest edges initially bonded to the substrates (Supplementary Figure S3).

Interestingly, the characteristic edge length of the VS$_2$ nanosheets (defined as the length of the longest edge for each half-hexagonal VS$_2$ flake) could be increased greatly from ~10 μm to ~40 μm by reducing the H$_2$ flow rate from 8 sccm to 2 sccm (Figure 1c). The evolution of the average edge length as a function of H$_2$ flow rate is plotted in Figure 1d, which clearly demonstrates that the size of the VS$_2$ nanosheets could be finely tuned by changing the composition and concentration of the carrier gas. Without H$_2$ in the carrier gas, no VS$_2$ products were observed on the SiO$_2$/Si substrates. The hydrogen flow is thus expected to have triggered and determined the reaction rate of vanadium chloride with sulfur, which consequently influences the growth rate and nucleation density of the VS$_2$ nanosheets (more discussion about the growth mechanism can be found in Supplementary Note 1.2).

The chemical composition of the VS$_2$ nanosheets was further determined by electron dispersive spectroscopy (EDS) (Figure 1e) and X-ray photoelectron spectroscopy (XPS) characterizations (Figure 1f and g). Both techniques revealed the presence of only sulfur and vanadium elements in the as-synthesized samples. The chemical composition derived from XPS results is based on the charge equilibration of $V^{3+}$, $V^{4+}$, and $S^{2-}$ ions, thus the presence of elemental sulfur in Figure 1f would not affect the accuracy of the final results. It is noteworthy that the V:S atomic ratio derived from both EDS and XPS data consistently exhibits a slightly nonstoichiometric value of ~1.1:2, indicating the existence of interstitial V atoms possibly within the vdW gaps (for simplicity, the sample is still denoted as VS$_2$ hereinafter). More details about the chemical composition determination with EDS and XPS, as well as the crystal structure characterization with spectral techniques (including Raman and X-ray diffraction) for the VS$_2$ nanosheets, are provided in Supplementary Note 2.1 and Figure S4, S5.

**Detailed characterizations of the VS$_2$ nanosheets.** The vertically or flatly aligned VS$_2$ nanosheets grown on SiO$_2$/Si could be transferred onto arbitrary substrates by the conventional PMMA-assisted transfer method[39] (Route 1 in Figure 2a, using NaOH as the etchant). Intriguingly, for those vertically grown VS$_2$, transference could also be implemented by simply pressing fresh SiO$_2$/Si (or any other flat substrate) facedown onto as-grown samples (Route 2 in Figure 2a). This polymer-free transfer method avoids the wet-chemical etching process, and is thus suitable for producing high-quality VS$_2$ samples with clean adlayer-substrate interfaces and few surface organic/inorganic contaminants.

Figure 2b shows an optical image of VS$_2$ nanosheets on a fresh SiO$_2$/Si substrate prepared using the second transfer method, with the uniform purple contrasts indicating their ultrathin nature. Scanning electron microscope (SEM) images of such VS$_2$ nanosheets (Figure 2c) typically exhibit randomly distributed wrinkles within the sheets, hinting at both the flexibility and ultrathin characteristic of the samples. As a direct measurement, atomic force microscopy (AFM) was utilized to reveal the exact thicknesses of the nanosheets, which were found to fall within a narrow range of 5–8 nm (corresponding to 8–14 layers). One height image of a 6.7-nm-thick VS$_2$ sheet is presented in Figure 2d, demonstrating its excellent thickness uniformity and quasi-2D feature.

The VS$_2$ nanosheets were also transferred onto highly oriented pyrolytic graphite (HOPG) that is suitable for Kelvin probe force microscopy (KPFM) characterization. Figures 2e and f exhibit the topography and surface potential images, respectively, of the VS$_2$/HOPG sample. It is evident that, the VS$_2$ nanosheets with varied thicknesses show nearly invariant surface potential differences with respect to HOPG in each scan line, which can be further visualized by the line profiles along the same cross-sectional cuts of the two images (Figure 1g). With the work function of HOPG known ($\Phi_{\text{HOPG}} = 4.65$ eV) and used as a reference, the work function of VS$_2$ was calculated as $\Phi_{\text{VS}_2} = \Phi_{\text{HOPG}} - e\Delta V_{\text{dc}} = 4.37$ eV, where $\Delta V_{\text{dc}}$ ~280 mV is the surface potential difference of the two. Notably, this work function is suitable for making good electrical contact with semiconducting MoS$_2$[40], as is discussed later in this work.

A diverse set of transmission electron microscopy (TEM) techniques were further utilized to resolve the crystal structure of the CVD-derived $VS_2$ nanosheets. Figure 3a is the low-magnification TEM image of a characteristic half-hexagonal $VS_2$ nanosheet transferred on a copper grid. Corresponding selected-area electron diffraction pattern is provided in Figure 3b, from which the (100) interplanar distance $d_{100}$ can be revealed to be ~0.28 nm, and thus the in-plane lattice constant $a_0 = d_{100}/\cos30° \approx 0.32$ nm. Both values are in good agreement with those of bulk-phase $VS_2$[37]. Intriguingly, exceptional satellite spots emerge in addition to the hexagonal diffraction pattern of $VS_2$, forming a $(1 \times \sqrt{3})$ superstructure as highlighted by the magenta rectangle. This superlattice periodicity is attributed to the ordering of some interstitial V atoms within the vdW gaps, as suggested by the EDS and XPS results in Figure 1e–g. Indeed, further high-resolution TEM (HRTEM) imaging demonstrates that multiphasic structures of diverse superlattices could coexist within a $30 \times 30$ nm$^2$ region (Supplementary Figure S7), which possibly implies an inhomogeneous distribution of the interstitial V atoms.

In particular, aberration-corrected high-angle annular dark-field scanning TEM (HAADF-STEM), characterized with more comprehensible contrast information than HRTEM, was utilized to further distinguish the occupancy and coordination states of both V and S atoms. Figure 3c is a representative HAADF-STEM image, where V atoms, with a higher atomic number than S, are better visualized and constitute a triangular pattern hosted by the hexagonally arranged S atoms. In the zoomed-in image color-coded by HAADF intensity (Figure 3d), the V and S atoms can be more clearly differentiated as yellow and green balls, respectively, with the pattern being further verified by the corresponding intensity line profile shown in Figure 3e. Such atomic arrangements are in perfect accordance with the 1T-phase atomic model (Figure 3f), in which V atoms are octahedrally coordinated by sulfur atoms, and adjacent $VS_2$ layers are atomically aligned in the vertical direction, forming well-defined atomic columns of V and S.

This interlayer stacking geometry also provides octahedral coordination sites (dashed circles in Figure 3f) potential for accommodating interstitial V atoms within the vdW gaps. Indeed, STEM imaging reveals such intercalation-induced superstructures (Figure 3g, h), in which alternating bright and dim atomic chains of V readily contribute to a $(1\times\sqrt{3})$ superlattice, as verified by the intensity line profiles in Figure 3i. This superlattice image agrees well with the SAED pattern captured on the VS$_2$ nanosheets, which suggests the segregation of the interstitial V atoms into ordered V-rich regions. Furthermore, our multi-slice simulations (Supplementary Figure S8) indicate that the superstructured region should be the self-intercalated V$_5$S$_8$ phase, which possesses the same $(1\times\sqrt{3})$ periodicity as the STEM image in the top-view atomic model shown in Figure 3j. Interestingly, considering the perfect lattice coherency across different phases (Supplementary Figure S9, and the identical $\Delta L$ in Figure 3e, i), it is evident that V intercalation did not cause obvious structural distortion in the VS$_2$ frameworks.

**Multifunctional electrode applications of the VS$_2$ nanosheets.** In an octahedral crystal field, the $d$ band usually splits into two sub-bands to accommodate remnant non-bonding $d$ electrons[20]. Theoretically, this yields a filling factor of 1/3 for the lower-energy $t_{2g}$ band in 1T-VS$_2$, and contributes to its metallic behavior (Figure 4a). For a 6.7-nm-thick VS$_2$ sample, our four-probe electrical measurements indeed revealed an ultralow resistivity of ~3 Ω μm and a monotonic increase of the resistance above 300 K (Figure 4b). In the low temperature range, however, the resistance-temperature ($R$-$T$) curve exhibits two peculiar extrema at ~106 K and ~25 K, which are likely to be associated with multiple transitions between ordered electronic phases. According to previous literature, the peak at ~106 K possibly reflects the latent structural instability of the material[30], while the inflection point at ~25 K is proposed to be associated with the magnetic ordering from paramagnetism to antiferromagnetism[41,42]. Interestingly, both of the two transition temperatures increase monotonically with increasing VS$_2$ thickness (black arrows in Figure 4c), and the insulating feature below 30 K vanishes, switching from an upturn to a downturn when the sheet thickness exceeds 8 nm. This phenomenologically illustrates a significant thickness dependence of the collective electronic states hosted by our CVD-VS$_2$ sample, although the intricate physics behind this requires further investigation.

Such metallic nanosheets with excellent electrical conductivities can potentially facilitate a number of applications as efficient electrode materials for 2D electronics and energy-related fields. In view of their similar atomic structures and complementary electronic properties, the VS$_2$ nanosheets were combined with monolayer (1L-)MoS$_2$ to construct all-MX$_2$ electronic devices. This was accomplished either by artificial stacking of individually grown VS$_2$ and MoS$_2$ layers (Figure 4, and Supplementary Figure S10), or through a two-step all-CVD growth process (growth details shown in Supplementary Figure S11). Ni/Au source and drain electrodes were then fabricated on two adjacent VS$_2$ nanosheets that were heterostructured with underlying 1L-MoS$_2$/SiO$_2$/Si for device performance evaluation (Figure 4d). A similar Ni/Au-contacted device was also constructed directly on the 1L-MoS$_2$ crystals for comparison (Figure 4g).

Output characteristics of the two bottom-gated field-effect transistors (FETs) (Figure 4e, h) show that the VS$_2$-contacted monolayer MoS$_2$ device has an on-state current of 3.8 µA µm$^{-1}$ at $V_{ds}$ = 10 V and $V_{gs}$ = 30 V, which is 4 times higher than that of the Ni/Au-contacted counterpart. By fitting the linear regions ($V_{gs}$ = 20–30 V) in the corresponding transfer curves (black solid lines in Figure 4f, i), the field-effect mobilities ($\mu_{FE}$) for the VS$_2$- and Ni/Au-contacted FETs were estimated to be 7.8 cm$^2$ V$^{-1}$ s$^{-1}$ and 2.1 cm$^2$ V$^{-1}$ s$^{-1}$, respectively. The mobility was calculated using the standard formula of $\mu_{FE} = (dI_{ds}/dV_g)[L/(WC_iV_{ds})]$, where $V_{ds}$ = 1 V, $L$ = 6.0 µm is the channel length, $W$ = 15.4 µm is the channel width, and $C_i$ = 3.8×10$^{-4}$ F m$^{-2}$ is the capacitance of the 90-nm-thick SiO$_2$ dielectric layer. Notably, both transistors manifest similar threshold gate voltages (~ -5 V), on/off ratios (~6×10$^6$) and subthreshold swings (~2 V dec$^{-1}$) (Supplementary Table S1). Nevertheless, the remarkable improvement in on-state current and field-effect mobility for VS$_2$-contacted MoS$_2$ devices could be reproducibly obtained, suggesting a much-reduced contact resistance ($R_{c,VS_2}$) across the VS$_2$/MoS$_2$ interface. That means, the CVD-synthesized VS$_2$ nanosheets could serve as ideal electrode materials for constructing monolayer MoS$_2$ devices.

Despite of the difficulty in direct determination of $R_{c,VS_2}$ using the transfer length method, the on-state $R_{c,VS_2}$ could be estimated to be less than 1/4 of that of Ni/Au contacts ($R_{c,Ni}$), since $R_{c,VS_2}/R_{c,Ni} < (2R_{c,VS_2} + $ $R_{ch})/(2R_{c,Ni} + R_{ch}) = I_{on,Ni}/I_{on,VS_2} \sim 1/4$, where $R_{ch}$ is the channel resistance and $I_{on,Ni\ (VS_2)}$ is the on-state current of the Ni (VS$_2$)-contacted MoS$_2$ FET. This dramatic improvement is comparable with those of phase-engineered metallic MoS$_2$[43] or MoTe$_2$ contacts[44]. The greatly improved device performance is attributed to the high-quality vdW interface which is free of Fermi level pinning effects[45], thus enabling efficient charge injection from the VS$_2$ electrode into the MoS$_2$ channel (more discussion can be found in Supplementary Figure S12). Superior to metastable 1T-MoS$_2$ and 1T'-MoTe$_2$ contacts, the 1T-VS$_2$ nanosheet contact is more thermodynamically stable, showing no performance degradation even after several months. Although further improvement in growing continuous VS$_2$ layers and/or controlling the location of individual VS$_2$ nanosheets has to be realized for its scalable applications, our electrical measurement results demonstrate that this metallic 2D material is suitable for serving as electrical contacts to 2D semiconductors such as monolayer MoS$_2$. Moreover, the versatility of CVD for synthesizing both MoS$_2$ and VS$_2$ components should also facilitate the realization of all-MX$_2$ electronics with clean vdW interfaces at the initial growth stage (preliminary results are provided in Supplementary Figure S13 to evaluate the possible electronic structure change of pre-synthesized MoS$_2$ after VS$_2$ deposition).

The VS$_2$ nanosheets have also manifested impressive performance as supercapacitor electrodes for energy-related applications. The capacitive behavior of VS$_2$ nanosheets can be understood to originate from their ultrahigh quantum capacitance[46], $C_Q = e^2(dn/dE_F)$, where $n$ is the carrier density, and $dn/dE_F$ is the DOS at the Fermi level. Specifically, the half-filled $t_{2g}$ band in VS$_2$ (Figure 5b) contributes to a rather high DOS at $E_F$, which is in stark contrast to semimetallic graphene with a near-zero DOS at the Dirac point. This, combined with the electrochemical intercalation process (Figure 5a), has led to a charge-storage mechanism that is more analogous to pseudocapacitors than electrical double-layer capacitors[47].

To practically achieve such devices, high-density VS$_2$ flakes (Figure 5c) grown at a shorter distance from the VCl$_3$ precursor were sonicated and dispersed in a mixed water/isopropanol solvent, and then deposited onto a glassy carbon (GC) electrode using Nafion binding agent[38]. Such VS$_2$/GC electrodes were tested in 0.5 M sulfate solutions under a standard three-electrode configuration, in which saturated calomel and graphite rod served as the reference and counter electrodes, respectively. The resulting cyclic voltammograms (CVs) in H$_2$SO$_4$ performed in the potential window from -0.3 V to 0.6 V versus NHE at various scan rates are shown in Figure 5d (electrochemical results in Na$_2$SO$_4$ and K$_2$SO$_4$ are provided in Supplementary Figure S14). The near-rectangular nature of the CVs indicates the capacitive behavior of the electrodes. Galvonastatic charge/discharge measurements performed at various currents in H$_2$SO$_4$ (Figure 5e) exhibit characteristic triangular curve shapes, also suggestive of the capacitive nature.

Notably, the specific capacitance derived by integrating the cathodic parts of the CV curves was revealed to be as high as 8.6×10$^2$ F g$^{-1}$ at a scan rate of 5 mV s$^{-1}$ (red curve in Figure 4d). Even at a much higher rate of 200 mV s$^{-1}$, the calculated capacitance remains at 2.0×10$^2$ F g$^{-1}$, a value comparable to that of high-performance supercapacitors based on graphene[48] and 1T-MoS$_2$ nanosheets[49]. Moreover, Ragone plot of the VS$_2$ electrode in H$_2$SO$_4$ (Figure 5f) reveals both high power density (>1 W cm$^{-3}$) and high energy density (>0.01 Wh cm$^{-3}$), a typical trait of pseudocapacitors. Note that the volumetric parameters were derived using the apparent density (~0.76 g cm$^{-3}$) of the VS$_2$ film (Supplementary Note 4.2). Nevertheless, these values are still among the highest reported to date for a variety of electrode materials, making the VS$_2$ nanosheets a competitive candidate for fabricating next-generation energy-storage devices.

Interestingly, the capacitive performance of the VS$_2$/GC electrodes was found to correlate strongly with the alkali metal ions used. In Figure 5g, CVs at 20 mV s$^{-1}$ in four different sulfate solutions show that moderate capacitances of ~2×10$^2$ F g$^{-1}$ can be achieved for Na$_2$SO$_4$ and K$_2$SO$_4$, whilst the capacitive behavior is almost absent in Li$_2$SO$_4$. This can be more clearly seen in the plots of capacitance (*C*) versus scan rate (*v*) in Figure 5h, where the capacitive performances at all scan rates follow the same sequence of H$_2$SO$_4$ > K$_2$SO$_4$ ≈ Na$_2$SO$_4$ ≫ Li$_2$SO$_4$. Notably, the distinct cation dependence combined with the $C \propto v^{-1/2}$ relationship (blue dashed line in Figure 5h) indicates that, this capacitive behavior is associated with a semi-infinite linear diffusion process[50], namely, intercalation of alkali metal ions into the vdW gap, rather than simply electrostatic adsorption on the VS$_2$ surface. This has been further corroborated by our electrochemical impedance spectroscopy measurements provided in Supplementary Figure S15. Moreover, the observed performance sequence is also in perfect accordance with the order of hydrated radii of the four cations[51] (inset in Figure 5h), and a cutoff radius, 3.58 Å (Na$^+$) < $r_c$ < 3.82 Å (Li$^+$), is anticipated to mark the upper limit of cation size for intercalation. Additionally, the VS$_2$/GC electrodes were subjected to cyclic charge/discharge tests in these aqueous solutions (Figure 5i) and retained capacitances in excess of 85% after 1000 cycles, indicating their excellent performance stability.

**Discussion**

In summary, we have demonstrated a facile CVD route for synthesizing VS$_2$ nanosheets with sub-10 nm thicknesses and domain sizes of tens of micrometers. This metallic 2D material represents the last but essential component in supplement to semimetallic graphene, semiconducting MoS$_2$, and insulating *h*-BN within the 2D family. The ultrahigh intrinsic carrier density in combination with the stoichiometric variability of the 2D crystal has provided great opportunities for exploring the relationship between the structural instability and the ordered electronic phases. Most importantly, the conductive nanosheets have manifested great promise for a host of applications, *i.e.*, serving as electrical contacts for monolayer MoS$_2$ that reduce the contact resistances by a factor of 1/4 with respect to Ni/Au electrodes, and as supercapacitor electrodes in aqueous electrolytes with specific capacitances as high as 8.6×10$^2$ F g$^{-1}$. These findings will pave the way for realizing functionality integration and diversification with the 2D building blocks, for which VS$_2$ nanosheets will also find a place.

**Methods**

**CVD growth of VS$_2$ nanosheets.** The CVD growth was implemented in a thermal split tube furnace (Lindberg/Blue M)

equipped with a 1-in.-diameter quartz tube. Three quartz boats containing 100 mg sulfur (Alfa Aesar, ≥ 99.5%), 20 mg $VCl_3$ (Alfa Aesar, ≥ 99%), and $SiO_2$/Si substrates, respectively, were loaded into the tube from upstream to downstream. Prior to heating, the furnace tube was purged with 500 sccm Ar for 10 min, and then flowed with 100 sccm Ar and 2–10 sccm $H_2$ to create a preferable growth atmosphere. A typical growth recipe was as follows: the evaporation and deposition zones were heated to 275 °C and 600 °C, respectively, within 25 min, kept at the setpoint temperature for 10 min, and then naturally cooled down to ~350 °C before the furnace was opened for rapid cooling.

**Characterization.** The $VS_2$ nanosheets were systematically characterized using optical microscopy (Nikon LV100ND), Raman spectroscopy (Horiba, LabRAM HR-800), X-ray diffraction (D/MAX-PC 2500), XPS (Kratos AXIS Supra/Ultra with monochromatic Al K$\alpha$ X-ray), AFM/KPFM (Bruker Dimension Icon), SEM (Hitachi S-4800; acceleration voltage of 1–5 kV), TEM (FEI Tecnai F20; acceleration voltage of 200 kV) equipped with an energy-dispersive spectrometer, and HAADF-STEM (JEOL JEM-ARM200F; acceleration voltage of 80 kV).

**Electrical measurement.** Methyl methacrylate (MMA) and PMMA were spin-coated on $VS_2$/$MoS_2$ samples and subjected to electron beam lithography (JEOL 6510 with Nanometer Pattern Generation System) to define the source and drain pattern. Metal contacts of 15 nm Ni/50 nm Au were then deposited on the sample using thermal deposition. Electrical performance of fabricated FETs was measured with the Lake Shore TTPX Probe Station and Agilent 4155C semiconductor parameter analyzer.

**Electrochemical measurement.** The electrochemical measurements were performed in a three-electrode system based on an electrochemical workstation (CHI660E), where $VS_2$ nanosheets, graphite rod, and saturated calomel electrodes served as the work, counter, and reference electrodes, respectively. 0.5 M $H_2SO_4$, $Li_2SO_4$, $Na_2SO_4$, and $K_2SO_4$ were used as the electrolyte solutions. Cyclic voltammetry data were collected in between -0.3 V and 0.6 V vs. NHE with scan rates ranging from 5 mV/s up to 200 mV/s. Operational potential ranges were chosen to avoid hydrogen evolution at low potentials and the oxidation of VS$_2$ at high potentials.

**Data availability.** The data that support the findings of this study are available from the corresponding author upon request.

**References**

1. Xu, K. *et al.* Ultrathin nanosheets of vanadium diselenide: a metallic two-dimensional material with ferromagnetic charge-density-wave behavior. *Angew. Chem. Int. Ed.* **52**, 10477–10481 (2013).

2. Zhu, X. *et al.* Signature of coexistence of superconductivity and ferromagnetism in two-dimensional NbSe$_2$ triggered by surface molecular adsorption. *Nat. Commun.* **7**, 11210 (2016).

3. Zhuang, H. L. & Hennig, R. G. Stability and magnetism of strongly correlated single-layer VS$_2$. *Phys. Rev. B* **93**, 054429 (2016).

4. Wilson, J. A., Di Salvo, F. J. & Mahajan, S. Charge-density waves and superlattices in the metallic layered transition metal dichalcogenides. *Adv. Phys.* **24**, 117–201 (1975).

5. Straub, T. *et al.* Charge-density-wave mechanism in 2H-NbSe$_2$: photoemission results. *Phys. Rev. Lett.* **82**, 4504–4507 (1999).

6. Weber, F. *et al.* Extended phonon collapse and the origin of the charge-density wave in 2H-NbSe$_2$. *Phys. Rev. Lett.* **107**, 107403 (2011).

7. Eichberger, M. *et al.* Snapshots of cooperative atomic motions in the optical suppression of charge density waves. *Nature* **468**, 799–802 (2010).

8. Yokoya, T. *et al.* Fermi surface sheet-dependent superconductivity in 2H-NbSe$_2$. *Science* **294**, 2518–2520 (2001).

9. Sipos, B. *et al.* From Mott state to superconductivity in 1T-TaS$_2$. *Nat. Mater.* **7**, 960–965 (2008).

10. Ang, R. *et al.* Real-space coexistence of the melted Mott state and superconductivity in Fe-substituted 1T-TaS$_2$. *Phys. Rev. Lett.* **109**, 176403 (2012).

11. Calandra, M., Mazin, I. I. & Mauri, F. Effect of dimensionality on the charge-density wave in few-layer 2H-NbSe$_2$. *Phys. Rev. B* **80**, 241108 (2009).

12. Yu, Y. *et al.* Gate-tunable phase transitions in thin flakes of 1T-TaS$_2$. *Nat. Nanotechnol.* **10**, 270–276 (2015).

13. Xi, X. *et al.* Strongly enhanced charge-density-wave order in monolayer NbSe$_2$. *Nat. Nanotechnol.* **10**, 765–769 (2015).

14. Ugeda, M. M. *et al.* Characterization of collective ground states in single-layer NbSe$_2$. *Nat. Phys.* **12**, 92–97 (2016).

15. Bae, S. *et al.* Roll-to-roll production of 30-inch graphene films for transparent electrodes. *Nat. Nanotechnol.* **5**, 574–578 (2010).

16. Bonaccorso, F. *et al.* Graphene, related two-dimensional crystals, and hybrid systems for energy conversion and storage. *Science* **347**, 1246501 (2015).

17. Novoselov, K. S. *et al.* A roadmap for graphene. *Nature* **490**, 192–200 (2012).

18. Zurutuza, A. & Marinelli, C. Challenges and opportunities in graphene commercialization. *Nat. Nanotechnol.* **9**, 730–734 (2014).

19. Wang, Q. H., Kalantar-Zadeh, K., Kis, A., Coleman, J. N. & Strano, M. S. Electronics and optoelectronics of two-dimensional transition metal dichalcogenides. *Nat. Nanotechnol.* **7**, 699–712 (2012).

20. Chhowalla, M. *et al.* The chemistry of two-dimensional layered transition metal dichalcogenide nanosheets. *Nat. Chem.* **5**, 263–275 (2013).

21. Naylor, C. H. *et al.* Monolayer single-crystal 1T'-$MoTe_2$ grown by chemical vapor deposition exhibits a weak antilocalization effect. *Nano Lett.* **16**, 4297–4304 (2016).

22. Zhou, L. *et al.* Large-area synthesis of high-quality uniform few-layer $MoTe_2$. *J. Am. Chem. Soc.* **137**, 11892–11895 (2015).

23. Keyshar, K. *et al.* Chemical vapor deposition of monolayer rhenium disulfide ($ReS_2$). *Adv. Mater.* **27**, 4640–4648 (2015).

24. Cui, F. *et al.* Tellurium-assisted epitaxial growth of large-area, highly crystalline $ReS_2$ atomic layers on mica substrate. *Adv. Mater.* **28**, 5019–5024 (2016).

25. Novoselov, K. S. *et al.* Two-dimensional atomic crystals. *Proc. Natl. Acad. Sci. USA* **102**, 10451–10453 (2005).

26. Li, L. J. *et al.* Controlling many-body states by the electric-field effect in a two-dimensional material. *Nature* **529**, 185–189 (2016).

27. Xu, C. *et al.* Large-area high-quality 2D ultrathin $Mo_2C$ superconducting crystals. *Nat. Mater.* **14**, 1135–1141 (2015).

28. Whangbo, M. H. & Canadell, E. Analogies between the concepts of molecular chemistry and solid-state physics concerning structural instabilities. Electronic origin of the structural modulations in layered transition metal dichalcogenides. *J. Am. Chem. Soc.* **114**, 9587–9600 (1992).

29. Mulazzi, M. *et al.* Absence of nesting in the charge-density-wave system 1T-$VS_2$ as seen by photoelectron spectroscopy. *Phys. Rev. B* **82**, 075130 (2010).

30. Gauzzi, A. *et al.* Possible phase separation and weak localization in the absence of a charge-density wave in single-phase 1T-$VS_2$. *Phys. Rev. B* **89**, 235125 (2014).

31. Zhang, H., Liu, L.-M. & Lau, W.-M. Dimension-dependent phase transition and magnetic properties of VS$_2$. *J. Mater. Chem. A* **1**, 10821–10828 (2013).

32. Ma, Y. *et al.* Evidence of the existence of magnetism in pristine VX$_2$ monolayers (X = S, Se) and their strain-induced tunable magnetic properties. *ACS Nano* **6**, 1695–1701 (2012).

33. Oka, Y., Kosuge, K. & Kachi, S. Order-disorder transition of the metal vacancies in the vanadium-sulfur system: an experimental study. *J. Solid State Chem.* **23**, 11–18 (1978).

34. Katsuta, H., McLellan, R. B. & Suzuki, K. Physcio-chemical properties of the non-stoichiometric VS$_2$ and V$_5$S$_8$ phases. *J. Phys. Chem. Solids* **40**, 1089–1091 (1979).

35. Nozaki, H., Ishizawa, Y., Saeki, M. & Nakahira, M. Electrical properties of V$_5$S$_8$ single crystals. *Phys. Lett. A* **54**, 29–30 (1975).

36. Nozaki, H. *et al.* Magnetic properties of V$_5$S$_8$ single crystals. *J. Phys. Chem. Solids* **39**, 851–858 (1978).

37. Feng, J. *et al.* Metallic few-layered VS$_2$ ultrathin nanosheets: high two-dimensional conductivity for in-plane supercapacitors. *J. Am. Chem. Soc.* **133**, 17832–17838 (2011).

38. Yuan, J. *et al.* Facile synthesis of single crystal vanadium disulfide nanosheets by chemical vapor deposition for efficient hydrogen evolution reaction. *Adv. Mater.* **27**, 5605–5609 (2015).

39. Jiao, L. *et al.* Creation of nanostructures with poly(methyl methacrylate)-mediated nanotransfer printing. *J. Am. Chem. Soc.* **130**, 12612–12613 (2008).

40. Das, S., Chen, H.-Y., Penumatcha, A. V. & Appenzeller, J. High performance multilayer MoS$_2$ transistors with scandium contacts. *Nano Lett.* **13**, 100–105 (2013).

41. Hardy, W. J. *et al.* Thickness-dependent and magnetic-field-driven suppression of antiferromagnetic order in thin V$_5$S$_8$ single crystals. *ACS Nano* **10**, 5941–5946 (2016).

42. Niu, J. *et al.* Anomalous Hall effect and magnetic orderings in nano-thick V$_5$S$_8$. Preprint at https://arxiv.org/abs/1611.00584 (2016).

43. Kappera, R. *et al.* Phase-engineered low-resistance contacts for ultrathin MoS$_2$ transistors. *Nat. Mater.* **13**, 1128–1134 (2014).

44. Cho, S. *et al.* Phase patterning for ohmic homojunction contact in MoTe$_2$. *Science* **349**, 625–628 (2015).

45. Chuang, H.-J. *et al.* Low-resistance 2D/2D ohmic contacts: a universal approach to high-performance WSe$_2$, MoS$_2$, and MoSe$_2$ transistors. *Nano Lett.* **16**, 1896–1902 (2016).

46. Zhu, J. *et al.* Defect-engineered graphene for high-energy- and high-power-density supercapacitor devices. *Adv. Mater.* **28**, 7185–7192 (2016).

47. Salanne, M. *et al.* Efficient storage mechanisms for building better supercapacitors. *Nat. Energy* **1**, 16070 (2016).

48. Stoller, M. D., Park, S., Zhu, Y., An, J. & Ruoff, R. S. Graphene-based ultracapacitors. *Nano Lett.* 2008, 8, 3498–3502.

49. Acerce, M., Voiry, D. & Chhowalla, M. Metallic 1T phase $MoS_2$ nanosheets as supercapacitor electrode materials. *Nat. Nanotechnol.* **10**, 313–318 (2015).

50. Augustyn, V. *et al.* High-rate electrochemical energy storage through $Li^+$ intercalation pseudocapacitance. *Nat. Mater.* **12**, 518–522 (2013).

51. Nightingale, E. R. Phenomenological theory of ion solvation. Effective radii of hydrated ions. *J. Phys. Chem.* **63**, 1381–1387 (1959).

## Acknowledgements

This work was financially supported by the Ministry of Science and Technology of China (Grants Nos. 2016YFA0200103, 2016YFA0300602, 2012CB921404, 2012CB933404, 2013CB932603, and 2014CB921002), the National Natural Science Foundation of China (Grants Nos. 51472008, 51290272, 51222201, 21201012, 51121091, 51072004, 51201069, 51522212, 51421002, and 11574005), the Strategic Priority Research Program of Chinese Academy of Sciences (Grant No. XDB07030200), the Beijing Municipal Science and Technology Planning Project (No. Z151100003315013), the Open Research Fund Program of the State Key Laboratory of Low-Dimensional Quantum Physics (No. KF201601), and the ENN Energy Research Institute.

## Author contributions

Y.Z. and Z.L. conceived and supervised the research project. Q.J. developed and conducted the CVD growth of $VS_2$ nanosheets, and prepared all the $VS_2$ and $VS_2$/$MoS_2$ samples for characterizations and measurements, with C.L., Z.Z., and Q.F.'s assistance. Y.G. and L.G. performed the HAADF-STEM characterization. Q.J., C.L., Z.Z., Q.F., Y.Z., and J.S. carried out the OM, XPS, SEM, AFM, KPFM, and TEM characterizations. J.W., J.N., L.L., and X.W. fabricated the Hall and FET devices and conducted the electrical measurements. Q.J. and C.L. prepared the $VS_2$ electrodes and performed the eletrochemical measurements. Q.J., Y.Z., and Z.L. co-wrote the manuscript and all the authors contributed to the critical discussion and revision of the final manuscript accordingly.

## Additional information

**Supplementary Information** accompanies this paper at http://www.nature.com/ naturecommunications

**Competing financial interests:** The authors declare no competing financial interests.

**Reprints and permission** information is available online at http://npg.nature.com/reprintsandpermissions/

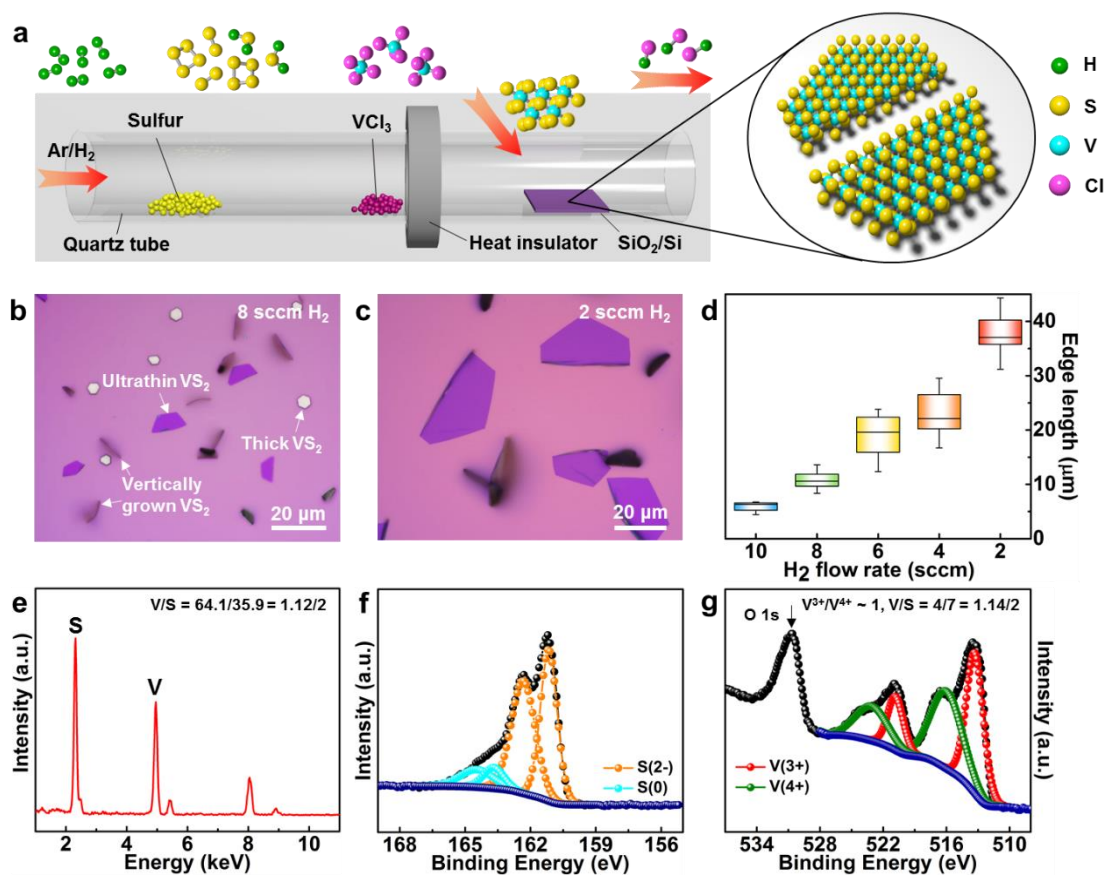

**Figure 1 | CVD growth and composition characterizations of VS$_2$ nanosheets.** (**a**) Schematic illustration of the CVD growth route. (**b**, **c**) Optical images of the VS$_2$ nanosheets grown under carrier gases of 100 sccm Ar mixed with 8 sccm and 2 sccm H$_2$, respectively. (**d**) Plot of VS$_2$ edge length versus H$_2$ flow rate. (**e**) EDS curve of a VS$_2$ nanosheet. The V:S atomic ratio is revealed to be 1.12:2 by calculating the integrated peak intensities of the two. (**f**) XPS curve of sulfur in the CVD-VS$_2$ sample. The curve is fitted with yellow and cyan doublet peaks, corresponding to anionic and elemental sulfur, respectively. (**g**) XPS curve of vanadium in the sample. The V$^{3+}$:V$^{4+}$ ratio derived from the fitting results is ~1:1, indicating a chemical formula of V$_4$S$_7$ (V$_2$S$_3$·2VS$_2$). The labeled oxidation peak at ~531 eV is found to originate from underlying SiO$_2$ rather than vanadium oxides (see Supplementary Figure S4 for more details).

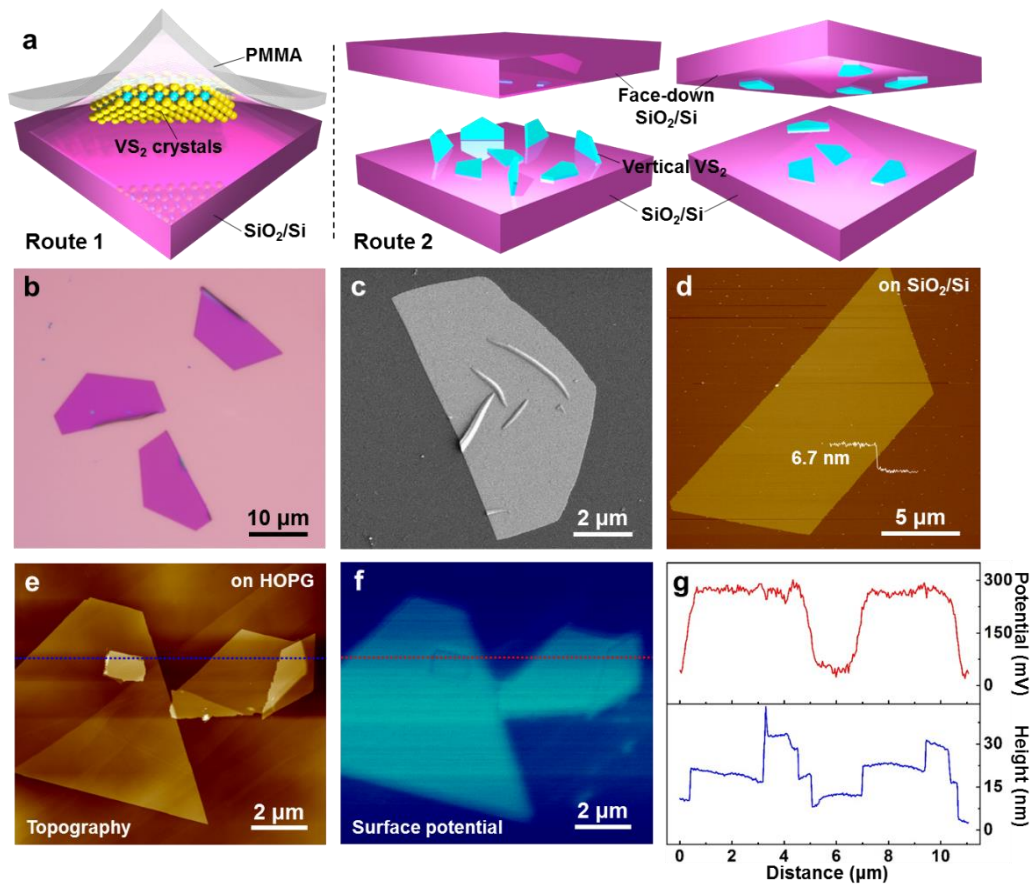

**Figure 2 | Transfer and microscopic characterizations of the VS$_2$ nanosheets.** (**a**) Schematic illustration of the two typical transfer methods for the VS$_2$ crystals. (**b**) Optical image of the half-hexagon-shaped VS$_2$ nanosheets. (**c**) SEM image of a VS$_2$ nanosheet as-grown on SiO$_2$/Si. (**d**) AFM height image of a VS$_2$ nanosheet. Inset is the corresponding height line profile. (**e**, **f**) Topography and surface potential images, respectively, of transferred VS$_2$ nanosheets on HOPG. The surface potential variations in the vertical direction within individual VS$_2$ nanosheets are derived from the unlevel background signal of HOPG (Supplementary Figure S6). (**g**) Profiles along the colored dash lines in **e** and **f**.

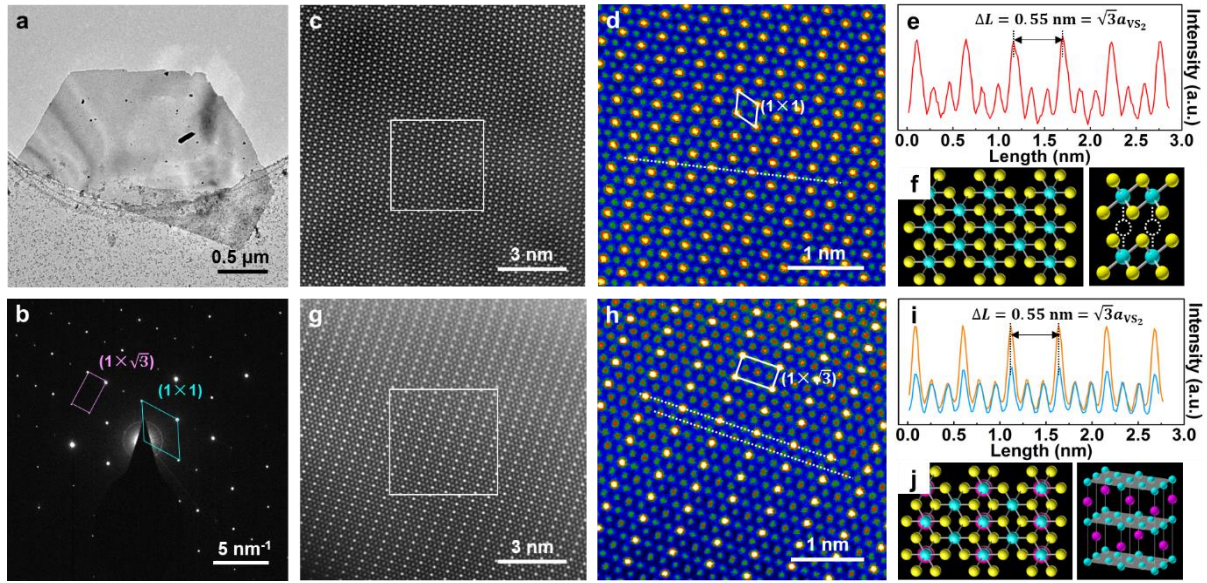

**Figure 3 | TEM characterization of the atomic structures of the VS$_2$ nanosheets.** (**a**) Low-magnification TEM image of a half-hexagonal VS$_2$ nanosheet. (**b**) SAED pattern captured within a 500 × 500 nm$^2$ area. (**c**) HAADF-STEM image of a pristine VS$_2$ region. (**d**) Zoomed-in image of the area highlighted by a white rectangle in **c**, with the false color coded according to ADF intensity. (**e**) Intensity profile along the white dashed line in **d**. (**f**) Top- and side-view images of the atomic model for 1T-VS$_2$. (**g**) HAADF-STEM image of a superstructured region. (**h**) Color-coded STEM image of the highlighted rectangle area in **g**. (**i**) Intensity profiles along the two dashed lines in **h**. (**j**) Atomic models for the self-intercalated V$_5$S$_8$ phase. Sulfur atoms in the 3D-view image were omitted for clear identification of intercalation sites (magenta balls).

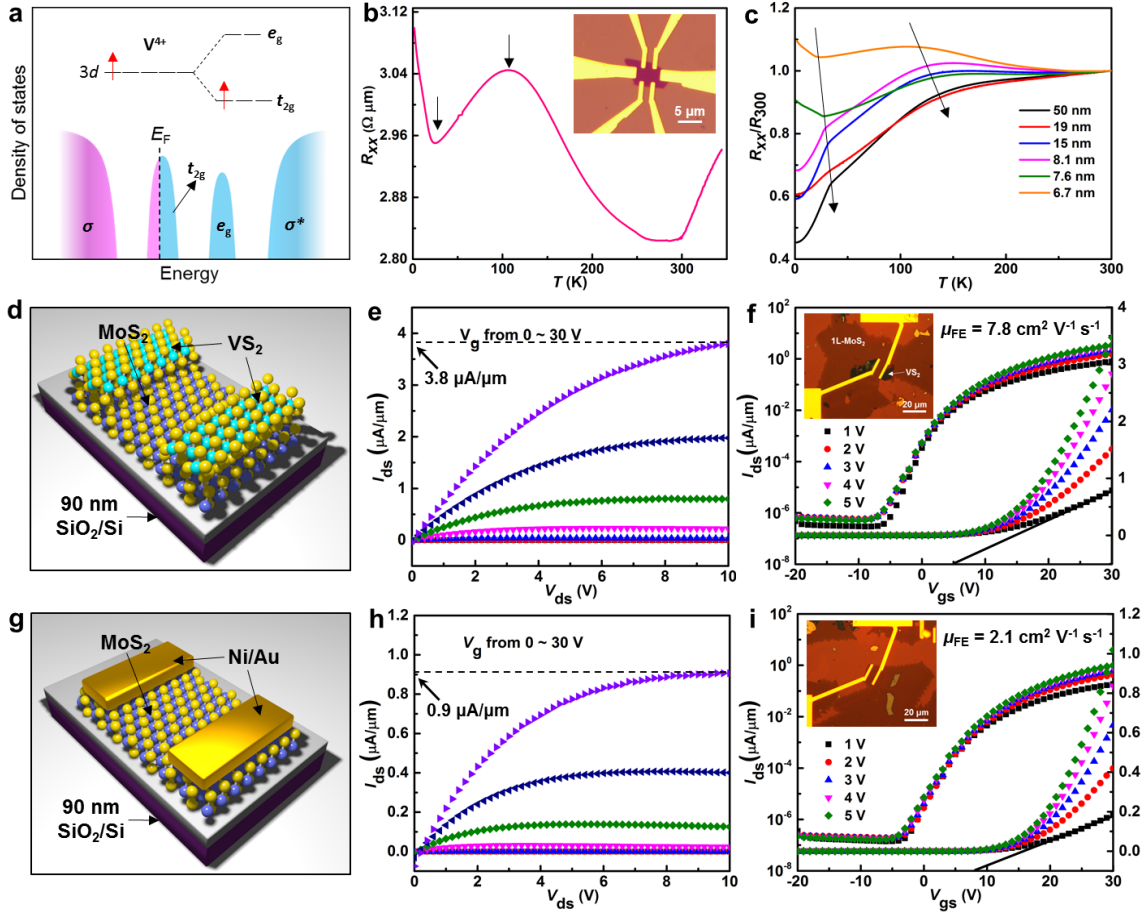

**Figure 4 | VS$_2$ nanosheets as the contact material for monolayer MoS$_2$ electronics.** (**a**) Schematic illustration of the DOS spectrum for 1T-VS$_2$. (**b**) Resistance-temperature curve of a 6.7-nm-thick VS$_2$ nanosheet. Inset is an optical image of a VS$_2$ Hall device. (**c**) Normalized *R-T* curves of VS$_2$ nanosheets with varied thicknesses. (**d**) Schematic model of VS$_2$-contacted monolayer (1L-) MoS$_2$ FET. (**e**, **f**) Output and transfer curves (linear and semilog plots) of VS$_2$-contacted 1L-MoS$_2$ FET. (**g**) Schematic model of Ni/Au-contacted 1L-MoS$_2$ FET. (**h**, **i**) Output and transfer curves of Ni/Au-contacted 1L-MoS$_2$ FET. Insets in **f** and **i** are corresponding optical images of the two devices.

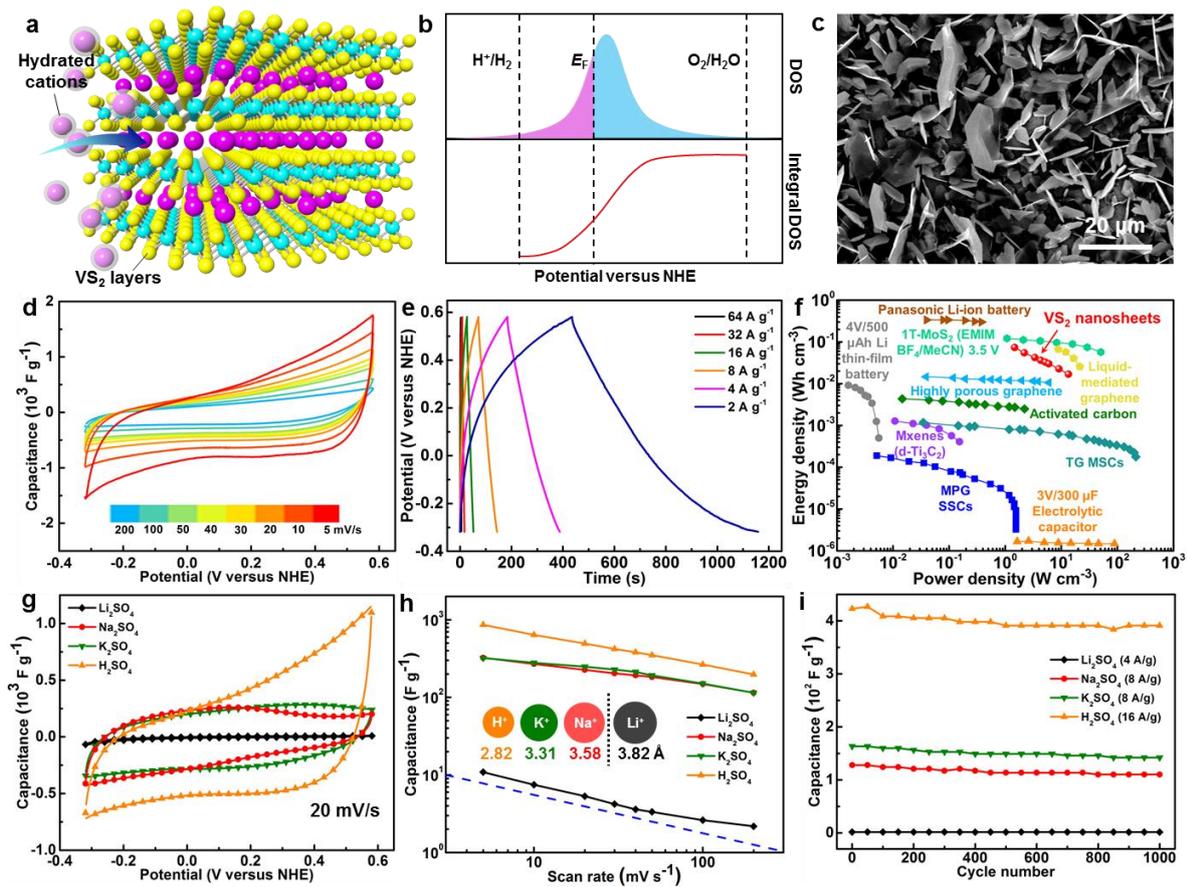

**Figure 5 | Electrochemical supercapacitor applications of CVD-VS$_2$ nanosheets.** (**a**, **b**) Schematic illustrations of the intercalation process and corresponding charge-storage mechanism for VS$_2$ nanosheets. (**c**) SEM image of the high-density VS$_2$ nanosheets. (**d**) CVs of VS$_2$/GC electrodes in 0.5 M H$_2$SO$_4$ at various scan rates. (**e**) Galvanostatic charge-discharge cycles from 2 A g$^{-1}$ to 64 A g$^{-1}$ in H$_2$SO$_4$. (**f**) Ragone plot of the VS$_2$ nanosheets in H$_2$SO$_4$ in comparison with various electrode materials[49]. The energy and power densities of VS$_2$ were calculated using the capacitances obtained in the three-electrode system. (**g**) CVs of VS$_2$/GC electrodes at 20 mV s$^{-1}$ in different 0.5 M aqueous sulfate solutions. (**h**) Scan rate dependence of the capacitance of VS$_2$/GC electrodes in sulfate solutions. (**i**) Capacitance retention over 1000 cycles in the four sulfate solutions.